\documentclass{article}%
\usepackage{amsfonts}
\usepackage{amsmath}
\usepackage{amssymb}
\usepackage{graphicx}%
\providecommand{\U}[1]{\protect\rule{.1in}{.1in}}
\begin{document}

\title{A geometric effect of quantum particles originated\ from the classicality of
their flow velocity}
\author{Tomer Shushi\\Center for Quantum Science and Technology\\\& Department of Business Administration,\\Guilford Glazer Faculty of Business and Management,\\Ben-Gurion University of the Negev, Beer-Sheva, Israel}
\maketitle

\begin{abstract}
In this short paper, we propose a new quantum effect that naturally emerges
from describing the quantum particle as a classical fluid. Following the
hydrodynamical formulation of quantum mechanics for a particle in a finite
convex region, we show how the maximum values of the wavefunction's amplitude
lie along the boundaries of the region when imposing a vanished quantum
potential, implying\ a classical flow velocity of the particle. The effect is
obtained for the case of particles in curved space, described by Riemannian
structures. We further show that such an effect\ cannot be achieved in the
relativistic regime when dealing with quantum particles in flat or curved spacetime.


\end{abstract}

\section{\bigskip Introduction}

Quantum mechanics is one of the most successful and profound theories in
modern science, offering an incredibly precise framework for understanding the
behavior of particles at the microscopic level. The Schr\"{o}dinger equation
stands as a cornerstone in quantum mechanics, providing a mathematical
description of the behavior of particles as wavefunctions. However, there are
different formulations of the quantum particle that provide different
interpretations about the nature of the quantum particles. Only a year after
Erwin Schr\"{o}dinger published his famous paper [1], which proposed the
Schr\"{o}dinger equation as the description for the dynamics of the quantum
particles, Erwin Madelung published another paper [2], showing that the
Schr\"{o}dinger equation can be converted into a dual problem that describes a
quantum fluid. Consider the Schr\"{o}dinger equation of a non-relativistic
particle with mass $m$ and some external potential $V,$ $i\hbar\partial
_{t}\psi=-\frac{\hbar^{2}}{2m}\nabla^{2}\psi+V\psi.$ Assuming that $\psi$ is a
smooth function, by taking the polar presentation $\psi\left(  \boldsymbol{x}%
,t\right)  =\sqrt{\rho\left(  \boldsymbol{x},t\right)  }e^{iS\left(
\boldsymbol{x},t\right)  /\hbar},$ we obtain the Madelung hydrodynamical
equations for the quantum particle,%
\begin{equation}
\partial_{t}\rho+\nabla\cdot\left(  \rho\boldsymbol{u}\right)  =0, \label{002}%
\end{equation}
and%
\begin{equation}
\partial_{t}\boldsymbol{u}+\boldsymbol{u}\cdot\nabla\boldsymbol{u}=-\frac
{1}{m}\nabla\left(  Q+V\right)  . \label{003}%
\end{equation}
Here $\rho$ is the density function of the particle, and $\boldsymbol{u=}%
\nabla S/m$ is the flow velocity of the particle, where (\ref{002}) is the
continuity equation of the fluid, and (\ref{003}) is the Hamilton-Jacobi
equation, with the addition of the quantum potential
\begin{equation}
Q\left(  \sqrt{\rho}\right)  =-{\frac{\hbar^{2}}{2m}}{\frac{\nabla^{2}%
\sqrt{\rho}}{\sqrt{\rho}}.} \label{004}%
\end{equation}
We thus see that by reformulating the Schr\"{o}dinger equation into a
hydrodynamical reformulation, another term pops up in the description of the
particle, which is the quantum potential. The quantum potential $Q$ gives the
coupling into the hydrodynamical equations (\ref{002})-(\ref{003}), in the
sense that the flow velocity $\boldsymbol{u}$ is coupled to the wavefunction's
amplitude through the quantum potential. We note that unlike the
Schr\"{o}dinger equation, here, the description of the particle as a fluid
excludes the notion of potentials since the equations of motion
describe\ forces instead of potentials. The external force is $F_{\text{ext}%
}=-\nabla V,$ while the quantum force is defined by $F_{\text{quant}}=-\nabla
Q.$

Since the introduction of quantum hydrodynamics by Madelung, a large amount of
literature has been developed in order to both study the foundations of
quantum mechanics and also to explore novel phenomena within quantum systems
[3-9], and\ in recent years, quantum hydrodynamics has been gaining more
attention with a growing amount of research in the field. When taking the
classical limit $\hbar\rightarrow0$ or the case of a massive particle
$m\rightarrow\infty,$ the quantum potential\ vanishes, and so (\ref{002}%
)-(\ref{003}) describes a classical fluid, where the flow velocity is
independent on the density function. Playing with the coefficient of $Q$ is a
trivial way to achieve this sort of classicality. However, we can achieve a
non-trivial classicality by imposing a vanished quantum force $F_{\text{quant}%
}=0,$ which boils down to the equation
\begin{equation}
Q\left(  \sqrt{\rho}\right)  =C\left(  t\right)  \label{005}%
\end{equation}
where $C$ is some real-valued function of time (see, [10]). This
framework\ shows us that for a suitable\ shape of the wavefunction, followed
by its density function $\rho,$ we can achieve a description of the particle
as a classical fluid.

In the following, we propose a geometric effect that emerges from\ quantum
particles with a vanished quantum force for particles in Riemannian structures
that describes curved space. We also show that this geometric effect does not
occur for particles in curved spacetime.

\section{Results}

We start with considering a non-relativistic quantum particle in a Riemannian
structure with the metric $ds^{2}=g_{ij}\left(  x\right)  dx^{i}dx^{j}.$ We
assume that our quantum particle exists in an open subset $M\subset%
%TCIMACRO{\U{211d} }%
%BeginExpansion
\mathbb{R}
%EndExpansion
^{n}$ of of the Riemannian space for $n>1$ spatial dimensions. The
Schr\"{o}dinger equation is then given by%
\begin{equation}
i\hbar\partial_{t}\psi=-\frac{\hbar^{2}}{2m}\nabla^{2}\psi+V\psi, \label{006}%
\end{equation}
where $\Delta=\nabla^{2}$ is the Laplace-Beltrami operator. The Hamiltonian
function for the Schr\"{o}dinger equation is (see, [5])
\begin{equation}
H\left(  \psi\right)  =\frac{\hbar^{2}}{2m}\left\Vert \nabla\psi\right\Vert
_{L^{2}}^{2}+\left\langle \left\langle V\psi,\psi\right\rangle \right\rangle
_{L^{2}}. \label{007}%
\end{equation}

Following the transformation $\psi\longmapsto\left(  \rho,S\right)  $,
$\psi:=\sqrt{\rho}e^{iS/\hbar}$, we can express $H\left(  \psi\right)  $ in
terms of $\left(  \rho,S\right)  .$ Following $\frac{\hbar^{2}}{2m}\left\Vert
\nabla\psi\right\Vert _{L^{2}}^{2}=\frac{\hbar^{2}}{2m}\left\langle
\left\langle \nabla\sqrt{\rho},\nabla\sqrt{\rho}\right\rangle \right\rangle
_{L^{2}}+\frac{1}{2m}\left\langle \left\langle \rho\nabla S,\nabla
S\right\rangle \right\rangle _{L^{2}},$ the Madelung Hamiltonian is given by
\begin{equation}
H\left(  \rho,S\right)  =\frac{1}{2m}\left\langle \left\langle \rho\nabla
S,\nabla S\right\rangle \right\rangle _{L^{2}}+\frac{\hbar^{2}}{2m}%
\left\langle \left\langle \nabla\sqrt{\rho},\nabla\sqrt{\rho}\right\rangle
\right\rangle _{L^{2}}+\left\langle \left\langle V\rho,1\right\rangle
\right\rangle _{L^{2}}. \label{010}%
\end{equation}

The Madelung equations are then obtained using Hamilton's equations (see,
again [5])%
\begin{equation}
\partial_{t}S=-\frac{\delta H\left(  \rho,S\right)  }{\delta\rho},\text{
\ }\partial_{t}\rho=\frac{\delta H\left(  \rho,S\right)  }{\delta S},
\label{011}%
\end{equation}
leading to%
\begin{equation}
\partial_{t}\boldsymbol{u}+\boldsymbol{u}\cdot\nabla\boldsymbol{u}+\frac{1}%
{m}\nabla\left(  Q_{g}+V\right)  =0, \label{012}%
\end{equation}
and%
\begin{equation}
\partial_{t}\rho+\operatorname{div}\left(  \rho\boldsymbol{u}\right)  =0,
\label{013}%
\end{equation}
where $\boldsymbol{u}:=\nabla S/m$ is the particle's flow velocity. The
quantum potential is then given by
\begin{equation}
Q_{g}\left(  \sqrt{\rho}\right)  =-\frac{\hbar^{2}}{2m}\frac{\Delta\sqrt{\rho
}}{\sqrt{\rho}}=-\frac{\hbar^{2}}{2m}\frac{\frac{1}{\sqrt{g}}\partial_{x^{i}%
}\left(  \sqrt{g}g^{ij}\partial_{x^{j}}\sqrt{\rho}\right)  }{\sqrt{\rho}},
\label{014}%
\end{equation}
where $\Delta$ is the Laplace-Beltrami operator, which also depends on the
metric $g.$ Similar to the case of Euclidean space, here, also, the quantum
potential can trivially vanish when taking the limits $\hbar\rightarrow0$ or
$m\rightarrow\infty.$

From the quantum potential $Q_{g}$ arises the quantum force
\begin{equation}
F_{Q_{g}}=-\nabla Q_{g}=\frac{\hbar^{2}}{2m}\frac{\nabla\frac{1}{\sqrt{g}%
}\partial_{x^{i}}\left(  \sqrt{g}g^{ij}\partial_{x^{j}}\sqrt{\rho}\right)
-\frac{1}{2\rho\sqrt{g}}\partial_{x^{i}}\left(  \sqrt{g}g^{ij}\partial_{x^{j}%
}\sqrt{\rho}\right)  \nabla\rho}{\sqrt{\rho}}.\label{015}%
\end{equation}
We produce a classical fluid by canceling the quantum force, with imposing
$F_{Q_{g}}=0,$%
\begin{equation}
Q_{g}=C\left(  t\right)  ,\label{016}%
\end{equation}
for some real-valued time-dependent function $C\left(  t\right)  $.\newline
Then, equation (\ref{016}) boils down to
\begin{equation}
g^{ij}\partial_{x^{i}}\partial_{x^{j}}P+\partial_{x^{i}}\left(  \sqrt{g}%
g^{ij}\right)  \partial_{x^{j}}P+\frac{2m}{\hbar^{2}}C\left(  t\right)
P=0.\label{017}%
\end{equation}
Recalling that $g^{ij}$ is symmetric and positive-definite, and assume that
$g^{ij}$ and $\partial_{x^{i}}\left(  \sqrt{g}g^{ij}\right)  \partial_{x^{j}%
}P$ have smooth (bounded) components, when setting $C\left(  t\right)
\equiv0,$ equation (\ref{016}) satisfies the strong maximum principle (smp).
The smp is a fundamental results in partial differential equations that states
that if a function attains its maximum within a bounded domain, it must either
be constant or touch the boundary. The smp guarantees that the maximum value
of the wavefunction's amplitude $P,$ as the solution of (\ref{017}), will be
on the boundary $\partial M$\ of the finite region the particle exists in, and
moreover, these maxima only exist on the boundary. For non-negative solutions
$P_{\text{class}}$ of (\ref{017}), this also implies that the density function
$\rho^{\ast}=P_{\text{class}}^{2}$ gains its maximum on the boundary, creating
a geometric effect of the particle which then moves along the boundary,
$\partial M$, of the region it exists in.

The following is an illustrative example of the proposed geometric effect in
the case of two spatial dimensions for a conformally flat curved space.

\bigskip%
%TCIMACRO{\FRAME{itbpF}{3.6382in}{1.2916in}{0in}{}{}{Figure}%
%maintain-aspect-ratio TRUE;  display "USEDEF";  valid_file "T";
%width 3.6382in;  height 1.2916in;  depth 0in;  original-width 8.4712in;
%original-height 2.989in;  cropleft "0";  croptop "1";  cropright "1";
%cropbottom "0";  tempfilename 'S9258K06.wmf';tempfile-properties "XPR";}} }%
%BeginExpansion
{\includegraphics[
height=1.2916in,
width=3.6382in
]%
{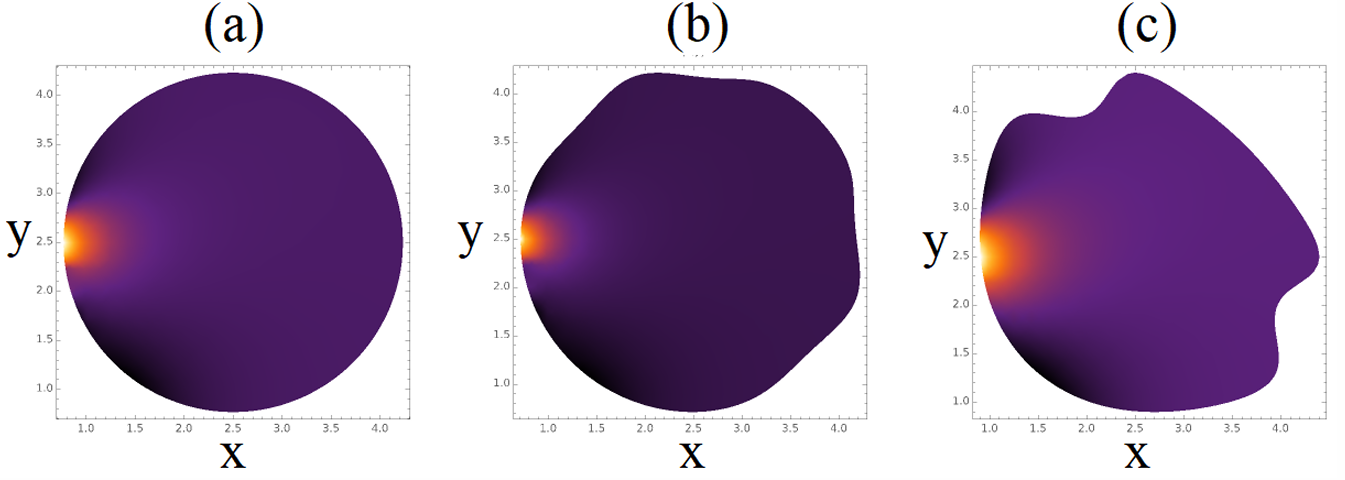}%
}
%EndExpansion

Figure 1. The density function $\rho^{\ast}\ $for a two-dimensional, $(x,y),$
quantum particle in a conformally curved space, with $g^{ij}=\Omega\left(
x,y\right)  \cdot diag\left(  1,1\right)  ,\mu=2.5,$\ the conformal factor
$\Omega\left(  x,y\right)  =\left(  e^{-\left(  x-\mu\right)  ^{2}%
}+e^{-\left(  y-\mu\right)  ^{2}}\right)  ,$ for a disc-shaped region (a) and
different convex-shaped regions (b)-(c).

\bigskip

As can be seen from Figure 1, $P_{\text{class}}$ is concentrated at the
boundaries of the convex closed regions. However, we note that the proposed
effect does not, in general, imply that $P_{\text{class}}$ will mainly exist
on the boundaries, and it can also flow into the closed region.

The desired wavefunction's amplitude, $P_{\text{class}},$ that gives the
geometric effect is, in general, not a solution of the Madelung equations
(\ref{012})-(\ref{013}). We, thus, have to find suitable quantum systems in
which (\ref{017}) can be satisfied. To do that, we consider the following
procedure (see [8]): At the first stage, we substitute the density function
$\rho^{\ast}=P_{\text{class}}^{2}$ corresponding to $P_{\text{class}}$ into
the continuity equation~(\ref{013}), $\partial_{t}\rho^{\ast}%
+\operatorname{div}\left(  \rho^{\ast}\boldsymbol{u}\right)  =0,$ which allows
us to find the flow velocity $\boldsymbol{u}^{\ast}$ in which this equation is
satisfied. The second stage is to substitute both $\rho^{\ast}$ and
$\boldsymbol{u}^{\ast}$ into the second Madelung equation (\ref{012}), which
then gives the desired external force $F_{\text{exp}}=-\nabla V$ that should
be imposed in order to have the desired shape of the wavefunction's amplitude,
with
\begin{equation}
F_{\text{exp}}=m\left(  \partial_{t}\boldsymbol{u^{\ast}+u^{\ast}}\cdot
\nabla\boldsymbol{u^{\ast}}\right)  +\nabla Q_{g}(P_{\text{class}%
}).\label{020}%
\end{equation}

In the hydrodynamical formulation of quantum mechanics, $\boldsymbol{J=}%
\rho\boldsymbol{u}$ is defined as the current density of the function, and so
the continuity equation~(\ref{013}), $\partial_{t}\rho+\operatorname{div}%
\left(  \boldsymbol{J}\right)  =0$ essentially describes the conservation of
probability $\rho$ in the system. We note that in case the current density
does not have turbulence behavior, which is manifested by a zero\ curl,
$\nabla\times\boldsymbol{J=0},$ and assuming that $\boldsymbol{J}$ goes to
zero at the limits $\left\vert x_{j}\right\vert \rightarrow\infty,$
$j=1,2,...,N,$ then we can write $\boldsymbol{J}$ as the divergence of a
scalar function $\phi,$ $\boldsymbol{J=}\nabla\phi.$ This means that we can
write the continuity equation as a Poisson's equation, $\Delta\phi
=-\partial_{t}\rho^{\ast}$. In the case of flat space, the solution takes an
explicit form, leading to the flow velocity $\boldsymbol{u=J}/\rho=\frac
{1}{P_{\text{class}}^{2}\left(  \boldsymbol{x},t\right)  }\nabla\int_{%
%TCIMACRO{\U{211d} }%
%BeginExpansion
\mathbb{R}
%EndExpansion
^{N}}\frac{\partial_{t}P_{\text{class}}^{2}\left(  \boldsymbol{x}^{\prime
},t\right)  }{4\pi\left\vert \boldsymbol{x}-\boldsymbol{x}^{\prime}\right\vert
}d\boldsymbol{x}^{\prime}.$

In the following, we show that, unlike the case of curved space, when we are
dealing with a quantum particle in curved spacetime, such a classicality of
the quantum particle does not bring the proposed geometric effect. This
feature is directly rooted in the definition of the spacetime metric
$g^{\mu\nu}.$

\subsection{Classicality of\ quantum particles in curved spacetime}

Consider a relativistic spinless quantum particle in $\left(  3+1\right)
$\ curved spacetime, modeled by\ the\ Klein-Gordon equation in curved
spacetime%
\begin{equation}
-g^{\mu\nu}\partial_{\mu}\partial_{\nu}\Psi+g^{\mu\nu}\Gamma_{\mu\nu}^{\sigma
}\partial_{\sigma}\Psi+\frac{m^{2}c^{2}}{\hbar^{2}}\Psi+U\Psi=0 \label{21}%
\end{equation}
where $x=\left(  x_{0}=ct,x_{1},x_{2},x_{3}\right)  ,$ and \bigskip
$\Gamma_{\mu\nu}^{\sigma}=\frac{1}{2}g^{\sigma\beta}\left(  \partial_{\mu
}g_{\beta\nu}+\partial_{\nu}g_{\beta\mu}-\partial_{\beta}g_{\mu\nu}\right)  $
is the christoffel symbol.

To achieve classicality as followed by quantum hydrodynamics, we consider the
polar representation, as before, $\Psi:=\sqrt{\rho}e^{iS/\hbar},$ with noting
that while $\sqrt{\rho}$ describes the wavefunction's amplitude, it does not
describe the density function of the quantum particle. Then, by substituting
the polar representation into the Klein-Gordon equation (\ref{21}) we
immediately obtain the corresponding Madelung equations
\begin{equation}
g^{\mu\nu}\partial_{\mu}S\cdot\partial_{\nu}S+2m_{0}Q_{g}\left(  P\right)
+m^{2}c^{2}+\hbar^{2}U=0, \label{22}%
\end{equation}
and%
\begin{equation}
\nabla_{\mu}J^{\mu}=0, \label{23}%
\end{equation}
with the quantum potential%
\begin{equation}
Q_{g}\left(  P\right)  =-\frac{\hbar^{2}}{2m_{0}}g^{\mu\nu}\frac{\partial
_{\mu}\partial_{\nu}P-\Gamma_{\mu\nu}^{\sigma}\partial_{\sigma}P}{P},
\label{24}%
\end{equation}
where $\rho=P^{2},$ and $J^{\mu}:=P^{2}g^{\mu\nu}\nabla_{\nu}S$ is the
four-current, and $m_{0}>0$ is a constant in units of mass.\ To achieve
classicality, we impose a vanished quantum force $F_{Q}=0$, which boils down
to
\begin{equation}
Q_{g}\left(  P\right)  +\lambda=0 \label{Q1}%
\end{equation}
\ for some real constant $\lambda$ (see, [6])$.$ Now, by setting
\[
\lambda=m_{eff}c^{2}%
\]
which is in units of energy, similar to the quantum potential, and taking
$m_{0}=m_{eff},$ for some effective mass $m_{eff},$ the equation (\ref{Q1})
can then be converted to%
\begin{equation}
-g^{\mu\nu}\partial_{\mu}\partial_{\nu}P+g^{\mu\nu}\Gamma_{\mu\nu}^{\sigma
}\partial_{\sigma}P+\frac{m_{eff}^{2}c^{2}}{\hbar^{2}}P=0. \label{26}%
\end{equation}
We, thus, achieve a vanished quantum force when the wavefunction's amplitude
follows the Klein-Gordon equation in empty curved spacetime with the same
metric as the original Klein-Gordon equation of the quantum particle, but with
some effective mass $m_{eff}$.

\bigskip Suppose now that our particle is given\ in a finite region $M\subset%
%TCIMACRO{\U{211d} }%
%BeginExpansion
\mathbb{R}
%EndExpansion
^{3+1}$. By setting $m_{eff}=0$, we have
\begin{equation}
g^{\mu\nu}\partial_{\mu}\partial_{\nu}P-g^{\mu\nu}\Gamma_{\mu\nu}^{\sigma
}\partial_{\sigma}P=0. \label{27}%
\end{equation}
While $g^{\mu\nu}$ is symmetric, which is one of the requirements for
obtaining the strong maximum principle, the components of $g^{\mu\nu}$ have
different signs, in general,
\begin{equation}
sgn\left(  g^{\mu\nu}\right)  \neq sgn\left(  g^{\mu\prime\nu^{\prime}%
}\right)  ,\text{ for }\mu\neq\mu^{\prime},\nu\neq\nu^{\prime}, \label{28}%
\end{equation}
and thus, the strong maximum principle cannot, in principle, be obtained for
such systems. The feature (\ref{28}) is fundamental in general relativity.\ In
fact, when (\ref{28}) is violated, we can have causality problems in the system.

\section{Discussion}

One of the most fundamental aspects of quantum mechanics is the boundary
between the quantum and classical behavior of the quantum particles. While at
the macroscopic level, quantum effects diminish and gradually give way to
classical mechanics, at the microscopic level, quantum mechanics governs the
behavior of the systems. The transition point from quantum to classical
behavior remains a subject of intense investigation and theoretical
exploration. The challenge lies in precisely defining when and how quantum
coherence dissipates, yielding to classical predictability. In this paper, we
have proposed a geometric effect for particles in Riemannian structures, which
naturally emerges when imposing the condition that their quantum potential
ultimately vanishes. We have shown a drastic difference between the
classicality of non-relativistic and relativistic systems, followed by the
proposed geometric effect. While in the non-relativistic regime, the geometric
effect holds, in the relativistic regime, such a geometric effect does not
hold, in principle, followed by the basic nature of the spacetime metric. We
propose to explore this in future research. We can extend the results into a
system containing $n>1$ coupled particles. Studying such systems can help get
new insights into the interplay between the proposed classicality with a
vanished quantum potential and the correspondence limit in which the system
containing many coupled particles transitions into a genuine macroscopic
classical system. We propose to explore it in future research.

\end{document}